\documentclass[12pt]{article}
\usepackage{sbc-template}
\usepackage{graphicx,url}
\usepackage[utf8]{inputenc}
\usepackage[english]{babel}
\usepackage{amsmath}  
\usepackage{booktabs}
\usepackage{graphicx}

\usepackage[acronym]{glossaries}

\newacronym{rbrs}{RBRS}{Review-based Recommender System}
\newacronym{rs}{RS}{Recommender System}
\newacronym{cf}{CF}{Collaborative Filtering}
\newacronym{mf}{MF}{Matrix Factorization}
\newacronym{nlp}{NLP}{Natural Language Processing}    
\newacronym{tfidf}{TF-IDF}{Term Frequency-Inverse Document Frequency}
\newacronym{bow}{BoW}{Bag-of-Words}
\newacronym{ai}{AI}{Artificial Intelligence}
\newacronym{ml}{ML}{Machine Learning}
\newacronym{mlp}{MLP}{Multi Layer Perceptron}
\newacronym{gat}{GAT}{Graph Attention Network}
\newacronym{nn}{NN}{Neural Network}
\newacronym{gcn}{GCN} {Graph Convolutional Network}
\newacronym{gnn}{GNN} {Graph Neural Network}
\newacronym{cnn}{CNN}{Convolutional Neural Network}
\newacronym{rnn}{RNN}{Recurrent Neural Network}
\newacronym{lda}{LDA}{Latent Dirichlet Allocation}
\newacronym{nmf}{NMF}{Non-Negative Matrix Factorization}
\newacronym{umap}{UMAP}{Uniform Manifold Approximation and Projection}
\newacronym{c-tf-idf}{c-TF-IDF}{class Term Frequency-Inverse Document Frequency}
\newacronym{hdbscan}{HDBSCAN}{Hierarchical Density-Based Spatial Clustering of Applications with Noise}
\newacronym{gte}{GTE}{thenlper/gte-small}
\newacronym{ncf}{NCF}{Neural Collaborative Filtering}
\newacronym{ngcf}{NGCF}{Neural Graph Collaborative Filtering}
\newacronym{svd}{SVD}{Singular Value Decomposition}
\newacronym{absa}{ABSA}{Aspect-based sentiment Analysis}
\newacronym{bert}{BERT}{Bidirectional Encoder Representations from Transformers}
\newacronym{roberta}{RoBERTa}{Robustly Optimized BERT Pretraining Approach}
\newacronym{rmse}{RMSE}{Root Mean Squared Error}
\newacronym{mae}{MAE}{Mean Absolute Error}
\newacronym{mse}{MSE} {Mean Squared Error}
\newacronym{map}{MAP}{Mean Average Precision}
\newacronym{mrr}{MRR}{Mean Reciprocal Rank}
\newacronym{gru}{GRU} {Gated Recurrent Unit}
\newacronym{lstm}{LSTM}{Long Short-Term Memory}
\newacronym{ndcg}{nDCG}{Normalized Discounted Cumulative Gain}
\newacronym{hr}{HR}{Hit Rate}
\newacronym{loo}{LOO}{Leave-One-Out}
\newacronym{cd}{CD}{Critical Difference}
\newacronym{hv}{HV}{Hypervolume}

\title{How Much Do Reviews Really Contribute? A  Study on Text-Enriched Matrix Factorization for Recommendations}

\author{Eduardo Ferreira da Silva\inst{1}, Mayki dos Santos Oliveira\inst{1}, Joel Machado Pires\inst{1} \\{Denis Dantas Boaventura}\inst{1}  Frederico Araújo Durão\inst{1}}

\address{Instituto de Computação -- Universidade Federal da Bahia (UFBA)\\
  Salvador -- BA -- Brazil
  \email{\{eduardoferreira, maykioliveira, joelpires,  denis.boaventura, fdurao\}@ufba.br}
}

\begin{document} 

\maketitle

\begin{abstract}
Incorporating textual reviews into a Recommender System has become a prominent strategy for enriching collaborative signals with semantic information. However, the actual contribution of review-derived representations remains an open question, particularly when strong collaborative baselines are employed. In this work, we systematically investigate the impact of textual information on Matrix Factorization by introducing and comparing three enrichment strategies over a common collaborative backbone. First, we propose a learnable gating mechanism that adaptively balances collaborative and textual signals during training. This mechanism is applied to two distinct review representations: (i) aggregated topic profiles extracted from user and item histories, and (ii) full text embedding representations derived from reviews. Additionally, we explore a cross-attention mechanism that identifies and emphasizes the most informative dimensions of the textual representation before fusion with collaborative factors. We evaluate six variants: pure, enriched with topic profiles and text via gating; enriched with topics and text via gating; and enhanced with cross-attention over textual features. Experiments across multiple review-based datasets reveal that although adaptive fusion mechanisms improve representation flexibility, the marginal contribution of textual signals remains limited compared to the collaborative backbone. These findings suggest that, under typical rating-prediction settings, collaborative information continues to dominate performance, raising important considerations for the effective integration of semantic review signals into recommendation models.
\end{abstract}
     


\section{Introduction}

Incorporating textual reviews into \gls{rs} has become a widely adopted strategy for enriching collaborative signals with semantic information~\cite{sentimentos_sbbd}. Reviews are often assumed to provide deeper insights into users’ preferences, revealing latent aspects and subjective opinions that numerical ratings alone cannot capture~\cite{Hasan2024}. Motivated by this premise, a growing body of research has proposed review-aware recommendation models that integrate textual representations via regularization, topic models, neural architectures, and attention mechanisms~\cite{sbbd_modelos_linguagem}. The underlying assumption is straightforward: if reviews explain why users assign ratings, then leveraging them should improve rating prediction and ranking performance.

However, recent investigations have questioned the consistency and robustness of these claims. In particular,~\cite{Sachdeva2020} highlights discrepancies in reported improvements across review-based \gls{rs} and emphasizes the need for careful empirical validation under controlled settings. Their findings suggest that textual signals do not uniformly outperform strong collaborative baselines, especially when experimental conditions deviate from narrowly defined scenarios. This reveals a broader and still unresolved issue regarding the effective contribution of textual reviews when integrated into well-optimized collaborative models.

Beyond inconsistencies in reported gains, the existing literature exhibits two closely related limitations. First, review-aware models are predominantly optimized and evaluated under regression objectives, relying on error-based metrics such as \gls{mae} and \gls{rmse}. While suitable for rating prediction, these metrics do not necessarily reflect ranking quality, the primary objective in many practical recommendation scenarios~\cite{sbbd_Calibrated_Rec}. Consequently, improvements in rating reconstruction may not translate into meaningful changes in top-N recommendation lists, leaving the impact of textual signals on ranking performance largely underexplored.

Second, despite the rich semantic content embedded in reviews, including aspects, sentiments, and contextual explanations, this content is typically integrated only to enhance rating prediction. The prevailing assumption is that better numerical reconstruction implies effective use of textual information. However, this perspective may be reductive, as it overlooks the broader potential of reviews to shape representation learning and influence interaction modeling beyond regression error. As a result, the true role and practical contribution of review text in collaborative recommendation remain insufficiently understood.

In this work, we revisit this issue from a controlled and systematic perspective. Rather than proposing an entirely new architecture, we adopt a strong \gls{mf} backbone and investigate multiple strategies for incorporating review information\footnote{Available at: \url{https://anonymous.4open.science/r/review_analysis-1DB7/}}. Matrix factorization is chosen for its well-established effectiveness and interpretability in modeling latent user–item interactions. We then enrich this backbone using three alternative mechanisms designed to fuse collaborative and textual signals: (i) a learnable gating mechanism applied to topic-based user and item profiles, (ii) a gating mechanism applied to full text embedding representations, and (iii) a cross-attention mechanism that identifies and emphasizes the most informative dimensions of textual representations before fusion.

\section{Related Works}
\label{sec:related_works}

The integration of review text into \gls{rs} has evolved from simple regularization strategies to complex neural architectures that treat textual signals as core sources of representation. Early approaches extended matrix factorization by incorporating topic models or textual regularizers to align latent factors with semantic dimensions extracted from reviews. Later, deep learning models began to encode textual content directly using convolutional or attention-based mechanisms.

DeepCoNN~\cite{zheng_deepconn_2017} is a foundational neural \gls{rbrs} that uses parallel convolutional networks to encode user and item reviews separately. It learns latent representations to predict ratings by extracting semantic features from text and combining them with collaborative signals. This approach shows that reviews can serve as primary inputs rather than just auxiliary constraints. Later developments incorporated attention mechanisms to improve performance and interpretability. For instance, DAML~\cite{liu_daml_2019} applies dual attention to selectively focus on informative review components and align them with rating features. Similarly, KANN~\cite{Liu2023} integrates knowledge entities extracted from reviews into a knowledge-aware attention framework, enabling explainable interactions between users and items at the semantic level. More recently, graph-based approaches such as LETTER~\cite{letter_Son2025} model user–user and item–item relations while preserving richer textual signals during propagation, thereby better capturing relational semantics.

Despite reported improvements in review-aware architectures, the empirical robustness of these gains has been questioned.~\cite{Sachdeva2020} conduct a comprehensive reassessment of review-based recommendation models and identify inconsistencies in experimental protocols and reported results. Their analysis shows that state-of-the-art review-aware methods often fail to outperform strong collaborative baselines in settings beyond narrowly defined ones. They emphasize the importance of rigorous evaluation and caution against overestimating the universal benefit of textual augmentation.  Complementarily,~\cite{Jurdi2021} analyzes the impact of natural noise in recommender datasets, highlighting how inconsistent user behavior can distort evaluation outcomes. They argue that variability in user profiles and rating inconsistencies significantly affect model performance and call for improved evaluation methodologies that account for data quality and behavioral coherence.

More recently, the emergence of Large Language Models (LLMs) has reshaped the discussion.~\cite{Tan2025} question whether specialized review-aware architectures remain necessary and introduce RAREval, a benchmark to systematically evaluate review contributions under zero-shot, few-shot, and fine-tuning settings. Their findings suggest that pretrained LLMs can implicitly capture semantic alignment and often outperform traditional review-based architectures, particularly in sparse and cold-start scenarios.

Unlike prior works that design increasingly complex neural or graph-based architectures to maximize the impact of review text, this research adopts a controlled perspective centered on matrix factorization. Rather than replacing collaborative filtering with deep textual encoders, we investigate how different mechanisms, such as gated fusion, cross-attention, and regularization, enrich a strong collaborative backbone. Additionally, motivated by the concerns raised in \cite{Sachdeva2020} and \cite{Jurdi2021}, our work emphasizes rigorous multi-metric evaluation and explicitly distinguishes between regression and ranking behavior. This allows us to analyze not only whether reviews improve prediction accuracy, but also whether they preserve collaborative transitivity and ranking effectiveness, providing a more balanced and systematic assessment of review integration.

\section{Matrix Factorization Backbone}

Despite the complexity of review-aware architectures, \gls{mf} remains a key element in many recommendation models. Often embedded in hybrid and neural systems, \gls{mf} efficiently captures latent user-item interactions through low-dimensional representations. In this work, \gls{mf} serves as the foundation for integrating textual enrichment mechanisms, allowing us to analyze how review information impacts regression error and ranking performance.

Formally, the \gls{mf} model is defined as follows. Given a user $u$ and an item $i$, the model learns latent factor vectors $\mathbf{p}_u, \mathbf{q}_i \in R^d$, where $\mathbf{p}_u$ and $\mathbf{q}_i$ denote the collaborative embeddings of user $u$ and item $i$, respectively and $d$ denotes the dimensionality of the latent space. To capture baseline popularity and bias effects, we include scalar bias terms $b_u$ (user bias), $b_i$ (item bias), and $b$ (global offset). The predicted interaction score $\hat{s}_{ui}$ is computed as:
\begin{equation}
\hat{s}_{ui} = \mathbf{p}_u^\top \mathbf{q}_i + b_u + b_i + b.
\end{equation}
To ensure the outputs are bounded and comparable to the ground truth, we apply a logistic sigmoid function $\sigma(\cdot)$ to map the scores into the specific rating range $[r{\min}, r{\max}]$:
\begin{equation}
\hat{r}_{ui} = \sigma(\hat{s}_{ui}) (r_{\max}-r_{\min}) + r_{\min}.
\end{equation}
The model is optimized using mean squared error (MSE) between $\hat{r}_{ui}$ and the observed rating $r_{ui}$.

\subsection{Gating Mechanism}

To integrate collaborative and textual representations, we employ a learnable gating mechanism that adaptively balances both information sources in the latent factor space \cite{graft}. Let $\mathbf{h}_u$ and $\mathbf{h}_i$ represent the raw topic-derived profiles extracted from review text. To ensure compatibility, these representations are aligned to the collaborative space through linear projections, yielding the textual embeddings $\mathbf{p}^{t}_u, \mathbf{q}^{t}_i \in R^d$:
\begin{equation}
\mathbf{p}^{t}_u = \mathbf{W}_u \, \mathbf{h}_u, 
\qquad
\mathbf{q}^{t}_i = \mathbf{W}_i \, \mathbf{h}_i,
\end{equation}
where $\mathbf{W}_u$, $\mathbf{W}_i$ are learnable projection matrices that map the topic features into the $d$-dimensional latent space.

Additionally, fusion is performed directly in the collaborative latent space using a gating network. For a generic entity $e \in \{u,i\}$ (user or item), the learned gate vector is computed as:
\begin{equation}
\mathbf{g}_e = 
\sigma\left(
\mathbf{W}_g 
\left[
\mathbf{z}^{cf}_e \, \Vert \, \mathbf{z}^{t}_e
\right]
+ \mathbf{b}_g
\right),
\end{equation}
where $\mathbf{z}^{cf}_e$ denotes the collaborative representation 
($\mathbf{p}^{cf}_u$ or $\mathbf{q}^{cf}_i$), 
$\mathbf{z}^{t}_e$ represents the textual/topic embedding 
($\mathbf{p}^{t}_u$ or $\mathbf{q}^{t}_i$), 
$\Vert$ denotes vector concatenation, 
$\sigma(\cdot)$ is the sigmoid activation ensuring 
$\mathbf{g}_e \in (0,1)^d$, 
and $\mathbf{W}_g$, $\mathbf{b}_g$ are learnable parameters.

The final fused representations are obtained via element-wise interpolation. 
For a generic entity $e \in \{u,i\}$ (user or item), we define:
\begin{equation}
\tilde{\mathbf{z}}_e 
= \mathbf{g}_e \odot \mathbf{z}^{cf}_e
+ (1 - \mathbf{g}_e) \odot \mathbf{z}^{t}_e,
\end{equation}
where $\odot$ denotes element-wise multiplication. 
By performing fusion in the aligned latent space, the model maintains compatibility with the matrix factorization objective while enabling adaptive integration of semantic information derived from reviews.

\subsection{Cross Attention}

In addition to gated interpolation, we investigate a cross-attention mechanism proposed in~\cite {attention_2017} to selectively integrate textual representations into the collaborative latent space. Unlike gating, which performs element-wise interpolation between two aligned vectors, cross-attention enables the collaborative embedding to dynamically attend to multiple textual components, emphasizing the most informative dimensions conditioned on the interaction.

Let $\mathbf{p}^{cf}_u \in \mathbf{R}^d$ denote the collaborative embedding of a user (analogously for items), and let 
$\mathbf{T}_u \in \mathbf{R}^{K \times d_t}$ represents a sequence of $K$ textual vectors derived from review representations (e.g., topic vectors or token/segment embeddings), where $d_t$ denotes the dimensionality of the textual space.

To enable interaction between both modalities, we project them into a shared latent space of dimension $d$:
\begin{align}
\mathbf{Q}_u = \mathbf{W}_Q \mathbf{p}^{cf}_u, \quad\quad
\mathbf{K}_u = \mathbf{T}_u \mathbf{W}_K, \quad\quad
\mathbf{V}_u = \mathbf{T}_u \mathbf{W}_V,
\end{align}
where $\mathbf{W}_Q \in \mathbf{R}^{d \times d}$ and 
$\mathbf{W}_K, \mathbf{W}_V \in \mathbf{R}^{d_t \times d}$ are learnable projection matrices. 
In this formulation, $\mathbf{Q}_u$ acts as the query, while $\mathbf{K}_u$ and $\mathbf{V}_u$ correspond to the key and value matrices derived from the textual sequence.
Attention scores are computed using scaled dot-product attention:
\begin{equation}
\mathbf{a}_u = 
\frac{\mathbf{Q}_u \mathbf{K}_u^\top}{\sqrt{d}},
\end{equation}
where the scaling factor $\sqrt{d}$ stabilizes gradients.
When a validity mask $\mathbf{m}_u \in \{0,1\}^{K}$ is available (e.g., to handle padded positions), invalid entries are assigned large negative values before normalization. The attention weights, context vector, and fused representation are then computed as:
\begin{align}
\boldsymbol{\alpha}_u &= 
\text{softmax}(\mathbf{a}_u), \quad\quad
\mathbf{c}_u = \boldsymbol{\alpha}_u \mathbf{V}_u, \quad\quad
\tilde{\mathbf{p}}_u = \mathbf{p}^{cf}_u + \mathbf{c}_u.
\end{align}

The vector $\mathbf{c}_u$ represents a weighted aggregation of textual information, where weights are determined by the relevance of each textual component to the collaborative query. The final representation $\tilde{\mathbf{p}}_u$ is obtained by adding residuals, preserving the original collaborative signal while allowing selective enhancement from review-derived features.


\subsection{Text Regularizer}

As a complementary strategy to adaptive fusion, we incorporate review information into matrix factorization via regularization. This approach preserves the pure collaborative interaction for rating estimation while encouraging alignment between collaborative and textual representations during training. In this context, textual information, either topic-based profiles or dense embedding representations, is integrated through an auxiliary regularization term. Let $\mathbf{p}^{t}_u$ and $\mathbf{q}^{t}_i$ denote the projected textual representations aligned to the collaborative latent space. The overall objective becomes:
\begin{equation}
\mathcal{L} = 
\mathcal{L}_{\text{MSE}} 
+ \lambda_u \|\mathbf{p}^{cf}_u - \mathbf{p}^{t}_u\|_2^2 
+ \lambda_i \|\mathbf{q}^{cf}_i - \mathbf{q}^{t}_i\|_2^2,
\end{equation}
where $\mathcal{L}_{\text{MSE}}$ is the rating prediction loss and $\lambda_u$, $\lambda_i$ control the influence of textual alignment.
This formulation treats textual information as a structural prior rather than a direct predictive component. The collaborative embeddings remain solely responsible for rating estimation, while textual representations guide their geometry in the latent space. As a result, the model allows us to analyze whether textual signals indirectly improve generalization, without explicitly altering the interaction function.

By contrasting this regularization-based approach with gated interpolation and cross-attention fusion, we obtain multiple perspectives on how review-derived information influences collaborative recommendation. This comparative framework enables a controlled investigation into whether textual signals meaningfully reshape latent representations or whether their contribution remains marginal when strong collaborative baselines are employed.

\section{Topics Extraction}

The topic extraction stage aims to identify fine-grained aspects of user experience expressed in reviews, such as emotional tone, immersion, and technical quality, that are not captured by structured metadata. We evaluated two sentence embedding models: \texttt{thenlper/gte-small}~\cite{li2023towards}, a lightweight transformer designed for efficiency and scalability; and \texttt{all-mpnet-base-v2}\footnote{Available at: https://huggingface.co/sentence-transformers/all-mpnet-base-v2}, a larger model that provides richer semantic representations at a higher computational cost. The overall process follows the BERTopic framework \cite{grootendorst2022}. High-dimensional document embeddings are first reduced using UMAP and then clustered with K-Means to group semantically related reviews, prioritizing stable and coherent clusters. Topic representation is derived using a KeyBERT-inspired keyword-extraction method aligned with document embeddings. To improve semantic clarity, domain-specific stopwords are removed, and n-grams up to trigrams are considered. The resulting pipeline produces interpretable topic labels, per-review topic distributions, and low-dimensional visualizations to support subsequent analysis and model integration.

\section{Experimental Settings}
\label{sec:experimental_settings}

\subsection{Evaluation Methodology}

To ensure unbiased evaluation, we use a 5-fold cross-validation strategy with a leave-one-out (LOO) protocol. For each user, one interaction is held out for testing, while the remaining interactions are used for training. This approach simulates realistic next-item recommendations, particularly in sparse datasets, and is repeated across five folds to reduce variance. 

For ranking-based metrics, we follow the protocol of sampling 99 negative items per user.
To identify the relevant items for each user, we normalize ratings using z-scores. Items with significantly negative normalized scores are treated as negative. This personalized normalization accounts for individual user rating bias and helps reduce label noise during ranking evaluation. These items correspond to unseen or non-relevant interactions and are combined with the held-out positive item to form a candidate set of 100 items~\cite{ma2025negativesamplingrecommendationsurvey}. The model’s ranking quality is evaluated using \gls{hr}, \gls{ndcg}, and \gls{mrr}. These metrics emphasize the position of the relevant item within the top-K recommendations.

To extract semantic information from reviews, we employ BERTopic, which leverages transformer-based embeddings, clustering, and class-based TF-IDF to generate topic distributions. Each review is represented as a topic vector, and user/item textual profiles are constructed by aggregating the topic distributions of their associated reviews. These topic-aware representations are used as structured semantic signals within the recommendation architecture.

\subsection{Datasets}
The integration of Amazon Movies and TV, IMDb, and Rotten Tomatoes enables the analysis of \glspl{rs} from multiple evaluative perspectives.  Table~\ref{tab:datasets} summarizes the statistical information of each dataset. 

\begin{table}[ht]
    \small
    \centering
    \caption{Comparative statistics of the datasets before and after preprocessing.}
    \begin{tabular}{l*{6}{c}}
        \toprule
        & \multicolumn{2}{c}{\textbf{Amazon Movies and TV}}  & \multicolumn{2}{c}{\textbf{IMDb}}  & \multicolumn{2}{c}{\textbf{Rotten Tomatoes}} \\
        \cmidrule(lr){2-3}\cmidrule(lr){4-5}\cmidrule(lr){6-7}
        \textbf{Metric} & \textbf{Raw} & \textbf{Processed}& \textbf{Raw} & \textbf{Processed}& \textbf{Raw} & \textbf{Processed} \\
        \midrule
        Reviews         & 17.3M     & 1.4M      & 932.4K    & 264.9K    & 1.1M      & 543.5K   \\
        Users           & 6.5M      & 76K       & 427K      & 7k        & 11K       & 2.6k     \\
        Items           & 747.7K    & 27.3K     & 1.1K      & 1.1K      & 17.7K     & 17.4K    \\
        Reviews/User    & 2.664     & 19.12     & 3.28      & 37.07     & 100.06    & 208.17   \\
        Reviews/Item    & 23.173    & 53.18     & 12.28     & 230.40    & 63.77     & 31.13    \\
        Sparsity        & 99.99\%   & 99.93\%   & 99.81\%   & 96.77\%   & 99.42\%   & 98.80\%  \\
        \bottomrule
    \end{tabular}
    \label{tab:datasets}
\end{table}

\subsection{Metrics}

To assess predictive accuracy in rating estimation, we use standard regression metrics from \gls{rs} literature \cite{ricci_recsys_handbook_2022}. We report \gls{mae} and \gls{rmse}, which measure the deviation between predicted and actual ratings in the test set. \gls{mae} offers a linear view of prediction errors, while RMSE applies a quadratic penalty, emphasizing larger errors and being more sensitive to outliers. 

\begin{equation}
    \resizebox{.9\textwidth}{!}{$s
    \begin{aligned}
        \text{MAE} = 
        \frac{1}{|\mathcal{T}|}
        \sum_{(u,i)\in\mathcal{T}}
        \left| r_{ui} - \hat{r}_{ui} \right|,
        \quad\quad
        \text{RMSE} =
        \sqrt{
        \frac{1}{|\mathcal{T}|}
        \sum_{(u,i)\in\mathcal{T}}
        \left( r_{ui} - \hat{r}_{ui} \right)^2
        },\quad\quad
        MRR = \frac{1}{|U|} \sum_{u \in U} \frac{1}{rank_u}.
    \end{aligned}
    $}
\end{equation}

To evaluate ranking performance under the \gls{loo} protocol, we use \gls{hr}, \gls{ndcg}, and \gls{mrr}, which assess the ordering quality of the top-$K$ recommendations. \gls{hr} verifies whether the relevant item appears within the top-$K$. \gls{ndcg} incorporates the item’s rank using a logarithmic discount, emphasizing higher positions. 
\begin{equation}
\resizebox{.9\textwidth}{!}{$
    \begin{aligned}
        HR@K = \frac{1}{|U|} \sum_{u \in U} \mathbf{I}(rank_u \leq K),
        \quad\quad
        nDCG@K = \frac{1}{|U|} \sum_{u \in U}
        \begin{cases}
        \dfrac{1}{\log_2(rank_u + 1)}, & \text{if } rank_u \leq K, \\
        0, & \text{otherwise},
        \end{cases}
    \end{aligned}
    $}
\end{equation}

\subsection{Hyperparameters analysis}

We performed a hyperparameter sensitivity study using Optuna\footnote{Available at: https://optuna.org/}, running 20 trials per dataset on a user-stratified 20\% subset to reduce computational cost while preserving the interaction structure. Each candidate configuration was trained for 15 epochs using the leave-one-out protocol (fold 0), with optimization guided by the validation \gls{rmse} loss. 
The search space covered dropout rates \{0.2, 0.5\}, learning rates \{1$e^{-4}$, 1$e^{-3}$\}, and four embeddings sizes \{256, 128, 64, 32\}, to assess the influence of model depth and capacity. The topic latent dimension was kept fixed to ensure fair comparison across trials. For each dataset, the final configuration was chosen based on validation \gls{rmse}. The optimal configuration identified across datasets utilized an embedding dimension of 64, a dropout rate of 0.2, and a learning rate of 0.001.

\section{Results}

Table \ref{tab:results_complete} reveals a clear distinction between prediction accuracy and ranking effectiveness across datasets. For the error metrics (\gls{rmse} and \gls{mae}), the hybrid architectures consistently outperform the Pure collaborative baseline. In particular, Gated Text achieves the best overall prediction performance on Amazon and IMDb, while Cross Attention attains the lowest error on Rotten Tomatoes. The margin of improvement over the Pure model is especially noticeable on Amazon (\gls{rmse} reduction of 0.085) and modest but consistent on Rotten Tomatoes (0.016). These results indicate that incorporating textual representations, especially when directly gated at the embedding level, effectively refines rating estimation. The small standard deviations across folds also suggest that these models perform stably on the regression task.

However, this improvement in rating prediction does not translate to superior ranking quality. Across all three datasets, the Pure model consistently achieves the best \gls{hr}@10, \gls{ndcg}@10, and \gls{mrr}@10. The gap is particularly large on Amazon, where the Pure model substantially outperforms the hybrids in ranking metrics. This pattern suggests that while textual signals help calibrate rating magnitudes, they may dilute the latent collaborative structure that drives effective item ordering. In contrast, the Pure model appears better at capturing global interaction patterns, interpreted as collaborative transitivity, thereby producing stronger top-K recommendations.

\begin{table*}[ht]
\centering
\caption{Performance comparison across datasets. Bold values indicate the best result for each metric within the corresponding dataset. The $\pm$ symbol denotes the standard deviation computed across folds.}
\label{tab:results_complete}
\resizebox{\textwidth}{!}{
\begin{tabular}{llcccccc}
\toprule
Model & Dataset & RMSE & MAE & HR@10 & NDCG@10 & MRR@10 \\
\midrule

Cross Attention & Amazon & 1.026 ($\pm$ 0.031) & 0.738 ($\pm$ 0.020) & 0.104 ($\pm$ 0.008) & 0.046 ($\pm$ 0.005) & 0.051 ($\pm$ 0.005) \\
Gated Text & Amazon &\textbf{0.985 ($\pm$ 0.032)} & \textbf{0.699 ($\pm$ 0.018)} & 0.135 ($\pm$ 0.009) & 0.062 ($\pm$ 0.004) & 0.064 ($\pm$ 0.003) \\
Gated Topic & Amazon & 1.043 ($\pm$ 0.037) & 0.770 ($\pm$ 0.029) & 0.205 ($\pm$ 0.015) & 0.114 ($\pm$ 0.012) & 0.107 ($\pm$ 0.011) \\
Pure & Amazon & 1.070 ($\pm$ 0.028) & 0.828 ($\pm$ 0.019) & \textbf{0.311 ($\pm$ 0.028)} & \textbf{0.187 ($\pm$ 0.021)} & \textbf{0.169 ($\pm$ 0.018)} \\
Text Regularizer & Amazon & 1.068 ($\pm$ 0.029) & 0.824 ($\pm$ 0.020) & 0.111 ($\pm$ 0.009) & 0.050 ($\pm$ 0.007) & 0.054 ($\pm$ 0.006) \\
Topic Regularizer & Amazon & 1.069 ($\pm$ 0.028) & 0.825 ($\pm$ 0.020) & 0.112 ($\pm$ 0.007) & 0.050 ($\pm$ 0.006) & 0.055 ($\pm$ 0.006) \\
\midrule
Cross Attention & IMDb & 3.311 ($\pm$ 0.012) & 2.948 ($\pm$ 0.013) & 0.200 ($\pm$ 0.052) & 0.095 ($\pm$ 0.010) & 0.085 ($\pm$ 0.009) \\
Gated Text & IMDb & \textbf{1.976 ($\pm$ 0.031)} & \textbf{1.448 ($\pm$ 0.030)} & 0.172 ($\pm$ 0.018) & 0.083 ($\pm$ 0.009) & 0.082 ($\pm$ 0.006) \\
Gated Topic & IMDb & 1.986 ($\pm$ 0.036) & 1.463 ($\pm$ 0.030) & 0.175 ($\pm$ 0.015) & 0.085 ($\pm$ 0.007) & 0.083 ($\pm$ 0.005) \\
Pure & IMDb & 1.988 ($\pm$ 0.028) & 1.467 ($\pm$ 0.027) & \textbf{0.210 ($\pm$ 0.015)} & \textbf{0.105 ($\pm$ 0.006)} & \textbf{0.099 ($\pm$ 0.004)} \\
Text Regularizer & IMDb & 1.988 ($\pm$ 0.031) & 1.468 ($\pm$ 0.031) & 0.114 ($\pm$ 0.025) & 0.048 ($\pm$ 0.013) & 0.051 ($\pm$ 0.014) \\
Topic Regularizer & IMDb & 1.990 ($\pm$ 0.032) & 1.468 ($\pm$ 0.032) & 0.114 ($\pm$ 0.039) & 0.048 ($\pm$ 0.019) & 0.050 ($\pm$ 0.013) \\
\midrule
Cross Attention & Rotten Tomatoes & \textbf{0.761 ($\pm$ 0.008)} & \textbf{0.598 ($\pm$ 0.007)} & 0.104 ($\pm$ 0.078) & 0.048 ($\pm$ 0.037) & 0.054 ($\pm$ 0.030) \\
Gated Text & Rotten Tomatoes & 0.777 ($\pm$ 0.007) & 0.610 ($\pm$ 0.004) & 0.256 ($\pm$ 0.016) & 0.128 ($\pm$ 0.009) & 0.116 ($\pm$ 0.008) \\
Gated Topic & Rotten Tomatoes & 0.775 ($\pm$ 0.005) & 0.611 ($\pm$ 0.003) & 0.272 ($\pm$ 0.023) & 0.143 ($\pm$ 0.015) & 0.130 ($\pm$ 0.012) \\
Pure & Rotten Tomatoes & 0.772 ($\pm$ 0.006) & 0.610 ($\pm$ 0.005) & \textbf{0.293 ($\pm$ 0.016)} & \textbf{0.149 ($\pm$ 0.009)} & \textbf{0.131 ($\pm$ 0.007)} \\
Text Regularizer & Rotten Tomatoes & 0.772 ($\pm$ 0.004) & 0.610 ($\pm$ 0.003) & 0.088 ($\pm$ 0.052) & 0.037 ($\pm$ 0.026) & 0.043 ($\pm$ 0.019) \\
Topic Regularizer & Rotten Tomatoes & 0.772 ($\pm$ 0.004) & 0.610 ($\pm$ 0.003) & 0.088 ($\pm$ 0.052) & 0.037 ($\pm$ 0.026) & 0.043 ($\pm$ 0.019) \\
\bottomrule
\end{tabular}
}
\end{table*}

Dataset characteristics also influence model behavior. The IMDb dataset exhibits a markedly different pattern: although Gated Text achieves the best error metrics, the Cross Attention model performs dramatically worse in \gls{rmse} and \gls{mae} (3.311), diverging from its behavior on the other datasets. This instability may be associated with IMDb’s substantially longer average review length. Longer and denser textual sequences increase model complexity and may introduce noise or overfitting in attention-based fusion mechanisms, particularly when textual variance is high. In contrast, shorter, more homogeneous reviews (as on Amazon and Rotten Tomatoes) appear more compatible with attention-based integration.

\subsection{Statistical Analysis}

To analyze the models’ behavior across five evaluation metrics and three datasets, we moved beyond isolated performance values and examined the consistency of their relative rankings. Rather than relying solely on absolute scores, the comparison emphasizes each method's stability across evaluation dimensions and data contexts, providing a more comprehensive assessment of overall robustness and competitiveness.
Following the non-parametric statistical procedure proposed by \cite{JMLRdemsar}, we adopted a global ranking framework combined with hypothesis testing to assess algorithmic superiority. 

To consolidate the metrics (\gls{rmse}, \gls{mae}, \gls{hr}, \gls{ndcg}, and \gls{mrr}) into a single comparative measure, we calculated the indicator \gls{hv}. Our evaluation protocol separates error and classification metrics to examine model behavior from different performance perspectives. For statistical comparison, we applied the \gls{hv} indicator in three configurations: (i) error metrics only, (ii) ranking metrics only, and (iii) all metrics together. This design allows us to assess whether improvements in ranking prediction translate into competitive ranking performance and to evaluate the model's overall balance.
This strategy transforms the multi-objective ranking vector into a single scalar value while preserving Pareto dominance relationships, enabling consistent, unbiased comparisons across models within the ranking scenario without conflating them with regression-based performance. 

For each dataset–metric context, algorithms were ranked from best to worst to avoid distortions arising from scale differences across domains. We then applied the Friedman test to verify whether ranking differences were statistically significant. Finally, global average ranks were computed, and \gls{cd} diagrams based on the Nemenyi post hoc test were used to identify statistically significant pairwise differences. This integrated procedure ensures that the selected model exhibits consistent, statistically supported performance across all evaluation dimensions.

\begin{figure}[ht]
    \centering
    \includegraphics[width=1\linewidth]{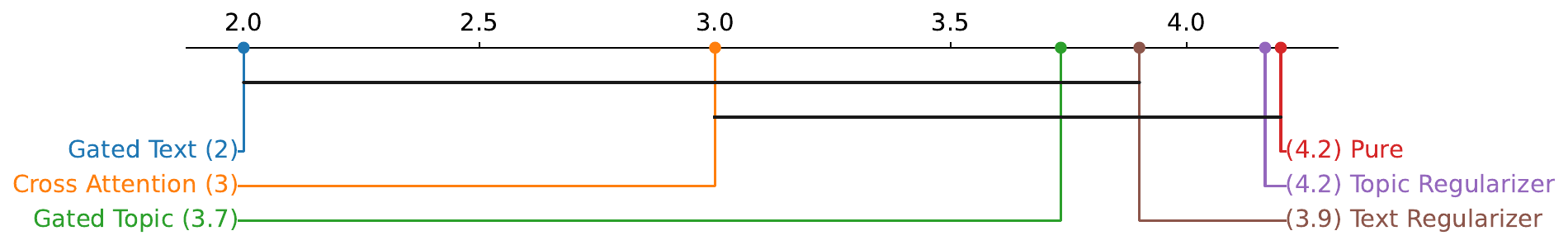}
    \caption{\gls{cd} diagram based on the Nemenyi post-hoc test, illustrating the average error metrics (\gls{rmse}, \gls{mae}) of the models across datasets. The models are ordered from left to right. Horizontal bars connect algorithms whose performance differences are not statistically significant at $\alpha$ = 0.05.}
    \label{fig:diagrama_dc_error}
\end{figure}

Figure~\ref{fig:diagrama_dc_error} presents the \gls{cd} diagram based on the average rankings across all datasets and folds, considering only the error metrics (\gls{rmse}, \gls{mae}). The Gated Text (2.0) method achieves the best average ranking (leftmost values), followed by Cross Attention (3.0) and Gated Topic (3.7), forming the main cluster. Although Gated Text achieves the best average ranking, its advantage over Cross Attention and Gated Topic is not consistently statistically significant, suggesting that the leading methods are competitive rather than exhibiting absolute dominance. 

\begin{figure}[ht]
    \centering
    \includegraphics[width=1\linewidth]{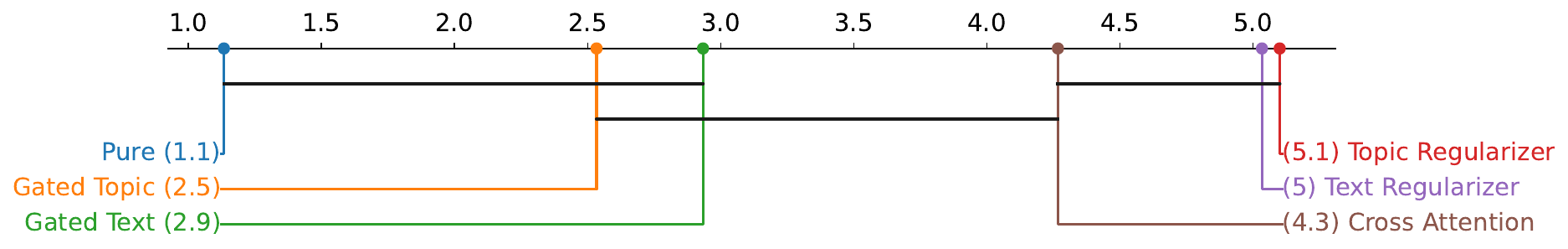}
    \caption{\gls{cd} diagram for the ranking metrics (\gls{hr}@10, \gls{ndcg}@10, and \gls{mrr}@10). Models are ordered by average rank (lower is better). Horizontal bars connect models whose performance differences are not statistically significant according to the Nemenyi post-hoc test.}
    \label{fig:diagrama_dc_rank}
\end{figure}

Figure~\ref{fig:diagrama_dc_rank} presents the calculated \gls{cd} diagram for the ranking metrics (\gls{hr}@10, \gls{ndcg}@10, and \gls{mrr}@10). In contrast to the behavior observed for the error metrics, the results are now reversed. The Pure model achieves the best (leftmost values) average ranking (1.1). The Gated Topic (2.5) and Gated Text (2.9) models form an intermediate group, while Cross Attention (4.3), Text Regularizer (5.0), and Topic Regularizer (5.1) occupy the weakest positions. The diagram indicates that Pure is statistically competitive with variants under tighter content constraints, but consistently ranks in the leftmost position, reinforcing its dominance in ranking performance. Models based on regularizers form a statistically indistinguishable cluster at the right end, reflecting their consistently inferior ranking behavior.

\begin{figure}
    \centering
    \includegraphics[width=1\linewidth]{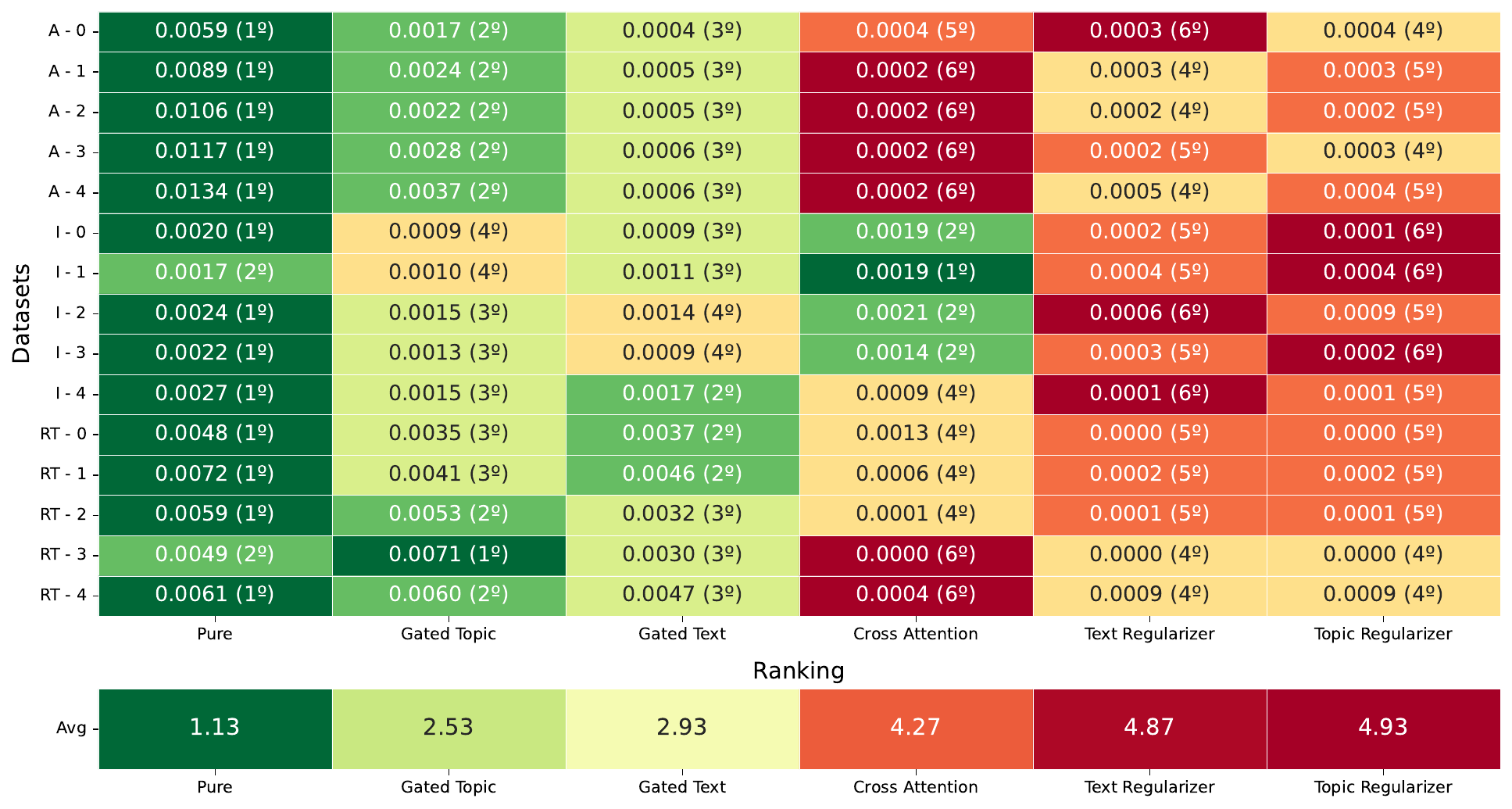}
    \caption{Ranking performance across datasets (A = Amazon, I = IMDb, and R = Rotten Tomatoes) and folds using the hypervolume (HV) indicator computed from all evaluation metrics. Each cell shows the HV value for a model in a dataset–fold combination, with its rank in parentheses (1º = best). The bottom row reports the average rank across all folds and datasets.}
    \label{fig:result_geral}
\end{figure}

\subsection{Discussion}

Incorporating review information into a matrix factorization framework effectively enhances recommendation quality. The results reveal a consistent pattern: textual augmentation improves rating prediction accuracy, but does not necessarily translate into superior ranking performance. Hybrid models, particularly the Gated Text variant, achieve lower \gls{rmse} and \gls{mae} across datasets, confirming that textual signals help refine regression estimates. This supports the idea that reviews provide complementary semantic information, thereby reducing prediction error.

\begin{figure}[ht]
    \centering
    \includegraphics[width=1\linewidth]{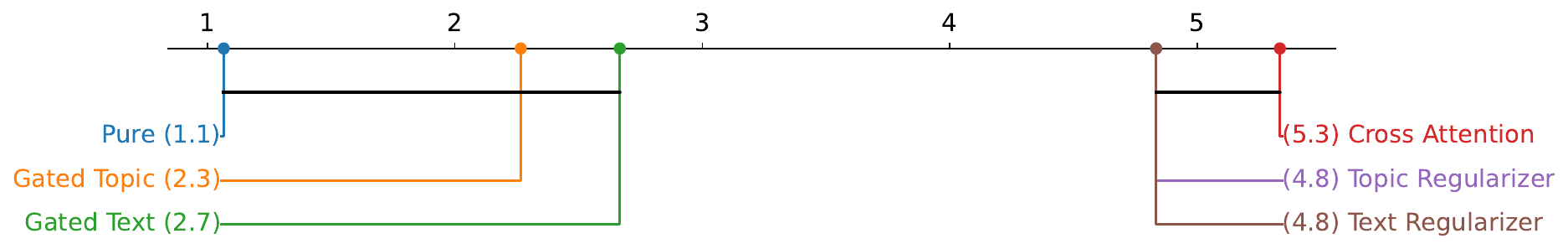}
    \caption{\gls{cd} diagram summarizing the overall comparison of models across all metrics. Horizontal bars connect groups of models whose differences are not statistically significant according to the Nemenyi post-hoc test.}
    \label{fig:cd_geral}
\end{figure}

In particular, although review-enhanced models introduce additional signals, they tend to reduce stability in the ranking metrics. As a result, the Pure model attains the best overall ranking across datasets and folds. This behavior is confirmed by the statistical comparison presented in Figure~\ref{fig:cd_geral}, which highlights the significant differences among model groups, while Figure~\ref{fig:result_geral} provides a detailed view of the results across folds, illustrating the consistent superiority of the Pure model in the aggregated ranking analysis.

However, when the evaluation perspective shifts to ranking metrics, the Pure collaborative model consistently outperforms the others. This indicates that collaborative signals remain more effective for capturing the relational structure necessary to order relevant items in top-$K$ recommendation tasks. One possible explanation is that emphasizing regression optimization may alter the latent space in ways that improve numerical rating estimation but weaken the transitive structure of user–item interactions. In other words, strengthening pointwise prediction may partially reduce the collaborative propagation effects that benefit ranking.

This behavior may also be interpreted through the lens of \textit{natural noise}. As discussed by \cite{Jurdi2021}, datasets inherently contain variability in user profiles. When reviews are incorporated, particularly in regression-oriented settings, models may become more sensitive to inconsistencies in user ratings. If user ratings do not align perfectly with textual sentiment or behavioral coherence, including textual signals may amplify this variability, improving local prediction fit without strengthening it and potentially diluting the global relational patterns required for ranking.

Therefore, the findings reinforce two important insights. First, improvements in regression metrics do not necessarily imply improvements in ranking quality. Second, collaborative structure alone remains highly robust for capturing global preference transitivity in top-$K$ recommendation scenarios. These results highlight the need to evaluate recommendation models across multiple metric families and suggest that future research should explicitly balance regression optimization, ranking effectiveness, and robustness to natural noise.

\subsection{Limitations and Improvement Points}

Despite the promising results, this study presents some limitations that open avenues for further improvement. First, the experimental analysis focuses exclusively on matrix factorization as the collaborative backbone. Although this choice ensures controlled comparisons and methodological clarity, it restricts the generalizability of the findings. 
Second, the notion of sentiment incorporated into the model was not formally defined theoretically. While sentiment signals were operationalized through extracted review representations, a clearer conceptual and mathematical definition of how sentiment is measured, scaled, and integrated into the recommendation process would strengthen interpretability and reproducibility. Providing a formal definition and detailing its computation would enhance transparency and methodological rigor.
Finally, the current approach does not leverage \gls{cnn} to capture localized, high-impact textual patterns in reviews.

\section{Conclusion and Future Works}

This work investigated three enrichment strategies for a hybrid matrix factorization model integrating review-based information. The empirical findings consistently indicate that hybrid variants tend to improve regression-oriented metrics, such as \gls{rmse} and \gls{mae}, while degrading ranking-oriented metrics. This pattern was observed across the Amazon, IMDb, and Rotten Tomatoes datasets, suggesting that the phenomenon is not dataset-specific. Although these results do not constitute definitive evidence, they provide strong empirical indications of a structural trade-off between error minimization and ranking effectiveness in review-aware hybrid models. 
Statistical hypothesis tests further support this interpretation by revealing two statistically distinct performance groups when models are evaluated under error-based versus ranking-based criteria. These findings reinforce the hypothesis that optimizing for rating prediction accuracy does not necessarily translate into improved item ordering, highlighting a fundamental tension between pointwise and ranking objectives in hybrid recommendation settings.

Future research should extend the proposed enrichment mechanisms beyond matrix factorization to other collaborative paradigms, including graph-based architectures and neural autoencoder frameworks. Exploring whether the observed trade-off persists under alternative modeling assumptions may provide deeper insights into the interaction between textual signals and collaborative representations. Additionally, investigating ranking-aware optimization strategies and more principled integration of textual information may help reconcile regression and ranking performance within unified recommendation frameworks.

\clearpage
\bibliographystyle{sbc}
\bibliography{sbc-template}

\end{document}